\documentclass[a4paper,fleqn]{cas-dc}

\usepackage[authoryear]{natbib}
\def\tsc#1{\csdef{#1}{\textsc{\lowercase{#1}}\xspace}}
\tsc{WGM}
\tsc{QE}
\begin{document}
\newcommand{\revb}[1]{{\color{blue} #1}}
\newcommand{\revf}[1]{{\color{Fuchsia} #1}}
\newcommand{\revm}[1]{{\color{Mahogany} #1}}
\newcommand{\revfg}[1]{{\color{ForestGreen} #1}}
\newcommand{\revbb}[1]{\textbf{{\color{blue} #1}}}

\let\WriteBookmarks\relax
\def\floatpagepagefraction{1}
\def\textpagefraction{.001}

\shorttitle{Review on XAI}    

\shortauthors{Cremades et al.}  

\title [mode = title]{Additive-feature-attribution methods: a review on explainable artificial intelligence for fluid dynamics and heat transfer}  




\author[1]{Andrés Cremades}[orcid=0000-0002-7052-4913]%[<options>]

\cormark[1]


\ead{andrescb@kth.se}


\credit{Investigation, Writing - Original Draft, Visualization}

\affiliation[1]{organization={FLOW, Engineering Mechanics, KTH Royal Institute of Technology},
            %addressline={}, 
            city={Stockholm},
%          citysep={}, % Uncomment if no comma needed between city and postcode
            postcode={SE-100 44}, 
            %state={},
            country={Sweden}}

\author[2]{Sergio Hoyas}[orcid=0000-0002-8458-7288]

% Footnote of the second author

% Email id of the second author
%\ead{}


\credit{Writing - Review \& Editing}

\affiliation[2]{organization={Instituto Universitario de Matemática Pura y Aplicada, Universitat Politècnica de València},
 %           addressline={}, 
            city={Valencia},
%          citysep={}, % Uncomment if no comma needed between city and postcode
            postcode={46022}, 
    %        state={},
            country={Spain}}

\author[1]{Ricardo Vinuesa}[orcid=0000-0001-6570-5499]%[<options>]

\cormark[1]

%\fnmark[3]

\ead{rvinuesa@mech.kth.se}

\credit{Writing - Review \& Editing, Project definition, Supervision}

\cortext[1]{Corresponding author}



\begin{abstract}
The use of data-driven methods in fluid mechanics has surged dramatically in recent years due to their capacity to adapt to the complex and multi-scale nature of turbulent flows, as well as to detect patterns in large-scale simulations or experimental tests. In order to interpret the relationships generated in the models during the training process, numerical attributions need to be assigned to the input features. One important example are the additive-feature-attribution methods. These explainability methods link the input features with the model prediction, providing an interpretation based on a linear formulation of the models. The SHapley Additive exPlanations (SHAP values) are formulated as the only possible interpretation that offers a unique solution for understanding the model. In this manuscript, the additive-feature-attribution methods are presented, showing four common implementations in the literature: kernel SHAP, tree SHAP, gradient SHAP, and deep SHAP. Then, the main applications of the additive-feature-attribution methods are introduced, dividing them into three main groups: turbulence modeling, fluid-mechanics fundamentals, and applied problems in fluid dynamics and heat transfer. This review shows thatexplainability techniques, and in particular additive-feature-attribution methods, are crucial for implementing interpretable and physics-compliant deep-learning models in the fluid-mechanics field.
\end{abstract}


\begin{highlights}
\item Review on the explainable artificial intelligence for fluid dynamics and heat transfer.
\item Description of the feature attribution methods used in fluid dynamics and heat transfer.
\item Classification of the main applications of explainable artificial intelligence for fluid dynamics and heat transfer.
\end{highlights}

%\nocite{*}

% Keywords

\maketitle

% Main text
\section{Introduction}

% how machine learning started
% how important it became
% where is it used and why-fluid mechanics
Since \citet{mcculloch1943} established the basis for artificial neural networks, proposing a mathematical formulation for the neural connections, data-driven methodologies have been developed, becoming an essential tool for modern research~\citep{jordan2015}. In the last years, machine learning has been widely used in different industrial~\citep{lee2016,bertolini2021} and scientific areas~\citep{mamoshina2016,albertsson2018} due to its ability to learn from data and to predict the solution for complex applications. These methods are exceptionally effective in solving those problems where the exact equations are not fully known or impossible to solve. Fluid mechanics is not an exception~\citep{brunton2021,vinuesa2024}, as in the fluid-mechanics field, non-linear relationships are generated within the flow as a consequence of the complexity of the Navier--Stokes equations~\citep{navier1827,Stokes2009} that govern it.

% What before ML
% What implied the ML
% Applications ML in fluids
Before applying machine learning, the analysis of turbulent flows was limited to large numerical simulations~\citep{ishihara2009,lee2013,alcantara2021} or experimental tests~\citep{laufer1975,talamelli2009,nagib2009}. Deep learning has opened new possibilities and horizons in fluid mechanics~\citep{vinuesa2023}, providing cost-efficient solutions for problems that in the past required long computation or experimental campaigns. Therefore, its use has increased exponentially in the last years for experimental~\citep{rabault2017,kim2024} and numerical~\citep{bar2019,wang2017,lee2020} studies. Concerning the former, machine learning exhibits the potential to improve measurement techniques, enhance experimental design, and control flows~\citep{vinuesa2023,yousif2023}. Regarding the latter, \citet{vinuesa2022} stated that machine learning has been used for developing reduced-order models~\citep{kaiser2014}, improving the turbulence models ~\citep{beck2019} and accelerating the simulations~\citep{li2020,kochkov2021}.

% Why do we need to understand the models
A wider understanding of turbulence transport~\citep{guastoni2021}, the effect of design parameters on the flow~\citep{du2022}, and how the operation conditions affect the performance~\citep{lin2020} are essential for future scientific discoveries and optimization of systems working with fluids, but they require a deeper explanation of the deep-learning models used for the predictions. However, these models are commonly used as black boxes and the relationships between the input parameters and the outputs are not analyzed.

% What has been done for understanding fluid mechanics-causality
% causality cannot explain deep-learning models
One of the most promising ideas used in the last years to understand turbulent flows is to analyze the causal relationships between the flow patterns and structures. \citet{osawa2024} applied interventional causality, modifying the flow inside a subdomain of the computational simulation, and then, calculating the evolution of the modified flow. This methodology transports the perturbations in time and evaluates their effect in the developed flow. As a result of this methodology, sweep-like structures~\citep{Lozano2012} were shown to be the ones affecting the flow the most. A different solution was proposed by \citet{lozano2022}, who used information theory to detect the causal relationships among various flow features~\citep{lozano2020,lozano2023}. Note that a deep-learning model generates a prediction from a set of input features, based on the data previously seen by the model. Therefore, to explain the patterns extracted by the model from the data it is necessary to analyze the impact of each input on the prediction.

% What do we need to understand DL models
% Game theory
% Adding explainable layer-implications-where has been used fields
In the case of deep-learning models, due to the large number of parameters and their complex architecture, a simplification is required to explain how each input feature affects the solution. Additive-feature-attribution methods have been proposed as a solution for explaining the internal relationships of the models by calculating a linear simplified representation. Each coefficient of the linear model assigns an importance attribution to the respective input feature. In order to calculate these attributions, recent research, such as the works by \citet{lundberg2017} or \citet{sundararajan2017}, have extended the game-theoretic methodology started by \citet{shapley1953} 70 years ago. These improvements allow adding an explainability layer to the deep-learning models. The use of eXplainable Artificial Intelligence (XAI), and specifically, additive-feature-attribution methods, provides human-interpretable results to the models and has been widely used in different fields, such as finances~\citep{mokhtari2019}, chemistry~\citep{wojtuch2021}, or process management~\citep{wang2022}.

% What have we done in the paper
The present work provides an overview of the methodologies that have been applied to improve the understanding of fluid mechanics by using XAI. First, the general framework of additive-feature-attribution methods is presented, focusing on those that have been applied to fluid mechanics. Then, the main applications of XAI in fluid mechanics are listed, starting with the improvement of turbulence modeling, continuing with the understanding of fluid dynamics, and finishing with miscellaneous industrial applications. The manuscript is concluded with a brief conclusion summarizing the application of XAI for fluid dynamics and providing a brief outlook of the future possibilities and impact of the additive-feature-attribution methods.

\section{Additive-feature-attribution methods}

As stated by \citet{lundberg2017}, the best explanation for a model is the model itself. In the previous statement,  explaining is defined as being able to establish textual or visual artifacts that provide qualitative understanding between the feature instances and the output of the model~\citep{ribeiro2016}. A clear example of this idea is Newton's second law: $F=ma$, where the acceleration $a$ of a body of mass $m$ is proportional to the force $F$. For this simple model, the equation is self-explanatory. An increase in the force will increase the acceleration proportionally  if the mass is kept constant. However, the interpretation of large models such as those in deep learning (DL) or complex physical equations, for instance, Navier--Stokes equations, is not straightforward. The main difference, from the explainability point of view, between Newton's second law and the Navier--Stokes equations remains in the linearity. Linear systems are simple to understand as modifications in the input generate proportional changes in the output.

The additive-feature-attribution methods, as proposed by \citet{lundberg2017}, exploit the explainable benefits of linear systems. The original model, $f$, is defined for a single input variable, $x$, which is calculated through a mapping function, $h$, from a simplified input, $x'$, $x=h\left(x'\right)$. Note that, in the present work, we focus on the local methods proposed by \citet{ribeiro2016}, which ensure that the model is explained using a simplified representation $g$, Equation (\ref{eq:additivefeatureattributionmethods}), which is equivalent to the original model, $g\left( z'\right)\approx f\left(h\left(z'\right)\right)$, when the simplified input $z'$, which contains a fraction of the non-zero features of the original simplified input $x'$ approximates it, $z'\approx x'$. 

\begin{equation}
    \label{eq:additivefeatureattributionmethods}
    g\left(z'\right)=\phi_0+\sum_{i=1}^{N}\phi_{i}z'_{i}\rm{.}
\end{equation}

\noindent The definition of the simplified model $g$ depends on a set of $N$ linear parameters $\phi_{i}$ known as SHAP (SHapley Additive exPlanations) values. These SHAP values express the effect of each single attribute $z'_{i}$ when they are present in the model. Indeed, $z'_{i}$ is a boolean parameter that takes the value $0$ if the attribute is absent in the model and $1$ if it is present.

The additive-feature-attribution methods provide a unique solution satisfying three main properties: local accuracy, missingness, and consistency. These properties are explained in more detail below:

\begin{itemize}
    \item Local accuracy: if $x=h\left(x'\right)$, then, the explainable model $g\left(x'\right)$ equals the system model $f\left(x\right)$, being $\phi_0$ the output of the explainable model when all the attributes are absent:
    \begin{equation}
        \label{eq:localaccuracy}
        f\left(x\right) = g\left(x'\right)=\phi_0+\sum_{i=1}^{N}\phi_{i}x'_{i}\rm{.}
    \end{equation}
    
    \item Missingness: the absent features should not have any effect on the output:
    \begin{equation}
        \label{eq:missingness}
        x'_{i} = 0 \to \phi_{i}=0\rm{.}
    \end{equation}

    \item Consistency: for two different models, $g$ and $g'$, the impact of the feature $i$ on the model $g'$ must be higher than in the model $g$ if the error of suppressing the feature $i$ in the model $g'$ is higher than in the model $g$:
    \begin{equation}
        \label{eq:consistency}
        \text{if } g'\left(z'\right)-g'\left(z'\backslash i\right) > g\left(z'\right)-g\left(z'\backslash i\right) \text{,}
    \end{equation}
    \begin{equation}
        \label{eq:consistency2}
        \text{then } \phi_{i}\left(g',z'\right)>\phi_{i}\left(g,z'\right)\text{.}
    \end{equation}
\end{itemize}

As demonstrated by \citet{young1985}, only the Shapley values proposed by \citet{shapley1953} satisfy at the same time the three previous properties. These Shapley values are defined by the expression below:

\begin{equation}
    \label{eq:shapley_values}
    \phi_{i} = \sum_{S\subseteq F\backslash\left\{i\right\}}\frac{\left|S\right|!\left(N-\left|S\right|-1\right)!}{N!}\left[f\left(S\cup i\right)-f\left(S\right)\right]\text{.}
\end{equation}

In equation (\ref{eq:shapley_values}), the Shapley value associated with the feature $i$, $\phi_{i}$, is calculated for the model $f$. These Shapley values weight the error of the prediction of the model between a subset, $S$ of the complete set of attributes, $F$, which does not contain the attribute $i$, and the subset when the feature $i$ is included: $\left[f\left(S\cup i\right)-f\left(S\right)\right]$. Note that the subset $S$ corresponds to the non-null values of the simplified input $z'$, or coalition. The weighting of this error is calculated by quantifying the probability of the feature to happen after the subset $S$, $\left(\left|S\right|!\left(N-\left|S\right|-1\right)!\right)/N!$, being $|S|$ the total number of present features of the subset and $N$ the total number of features of the model. The Shapley values of any player can be understood as the marginal contribution of the feature $i$ to the existing coalition $S$~\citep{myerson1991}.

As stated by~\citet{lundberg2017}, the exact calculation of the Shapley values is challenging, as the computational cost grows exponentially with the number of features, 
$2^{\rm{N}}$~\citep{jia2019}, where the number of parameters might reach $N\thicksim\mathcal{O}\left(10^9\right) \text{ to } \mathcal{O}\left(10^{10}\right)$ for the larger simulations of turbulent flows~\citep{iwamoto2004,iwamoto2005,hoy06,alcantara2018}. To overcome this limitation, different approximations, such as kernel SHAP~\citep{lundberg2017}, deep SHAP~\citep{lundberg2017}, tree SHAP~\citep{lundberg2018}, and gradient SHAP~\citep{erion2021} have been proposed in the literature.

\subsection{Kernel SHAP}

The Kernel SHAP is an approximation of the Shapley values which calculates the effect of each attribute by combining the previously discussed SHAP values with LIME (local interpretable model-agnostic explanations)~\citep{ribeiro2016}. LIME interprets the individual predictions of the model by approximating them locally. The objective function $\mathcal{L}$ is minimized as can be observed in Equation~(\ref{eq:lime}).  Indeed, the contribution to the output of each input feature, $\xi$, is calculated by minimizing the previous loss function, $\mathcal{L}$, which depends on the explanation model, $g$, defined in equation (\ref{eq:additivefeatureattributionmethods}), the original model, $f$, and a local kernel, $\pi_x$. Then, a penalization over the complexity of the model is included by the term $\Omega(g)$.

{\begin{equation}\label{eq:lime}
    \xi = \arg \min_{g \in \mathcal{G}}{\mathcal{L} (f,g,\pi_x)+\Omega(g)}.
\end{equation}}

Although the LIME expression present in Equation (\ref{eq:lime}) differs from the Shapley definition, Equation (\ref{eq:shapley_values}), it might satisfy local accuracy, missingness, and consistency conditions that the additive-feature-attribution method requires. For a heuristical definition of the loss function, $\mathcal{L}$, the local kernel $\pi_x$ and the regularization term, $\Omega(g)$, the LIME cannot adhere to the definition of the additive-feature-attribution methods. However, the local explanation of the LIME may satisfy the properties of the  methods, equation (\ref{eq:additivefeatureattributionmethods}), with the correct definition of the loss function, $\mathcal{L}$, local kernel $\pi_x$ and regularization term, $\Omega(g)$. For the kernel-SHAP framework, these functions are defined as follows:

 \begin{equation}\label{eq:complexitypenalty}
    \Omega(g) = 0,
\end{equation}
 \begin{equation}\label{eq:lossshap}
\mathcal{L}(f,g,\pi_x) = \sum_{z'\in Z}\left[f(h(z'))-g(z')\right]^2\pi_x(z'),
\end{equation}
 \begin{equation}\label{eq:kernelShap}
\pi_x(z') = \frac{N-1}{\left(\begin{array}{c}
N\\
\left| z' \right|\\
\end{array}\right) \left| z' \right| \left(N - \left| z' \right|\right)}\text{,}
\end{equation}

\noindent which are the only possible solutions of equation (\ref{eq:lime}) that satisfy the local accuracy, missingness, and consistency properties as demonstrated by \citet{lundberg2017}. In the previous equations the parameter $\left| z' \right|$ is the total number of non-zero structures present in the subset $z'$, and $Z$ is the total set of features.

The definition of the model in equation (\ref{eq:additivefeatureattributionmethods}) assumes a linear form, which differs from the loss function $\mathcal{L}$ defined as squared loss. However, the loss function can still be solved using a linear regression. In other words, the Shapley values can be computed using a weighted linear regression~\citep{charnes1988}. In addition, kernel SHAP reduces the computational cost of the Shapley values as they are approximated by computing a set of random samples.

Note that the simplified input $z'$ accounts for the present and absent features in the calculation. However, for most of the models, such as neural networks, an input feature cannot be absent. Therefore, these absent features are set to a non-informative value or reference value. In other words, the mapping function $h$ provides the input value $x$ for those features that are present and the reference value $x_r$ for the absent ones.

\subsection{Tree SHAP}

The previously explained exact Shapley values require an exponential computational cost as they need to evaluate a total of $2^N$ predictions. However, the tree-SHAP algorithm~\citep{lundberg2018} exploits the architecture of the tree models, which obtain the predictions by recursively partitioning the data space and fitting the model within each partition~\citep{loh2011}, and keeps track of the proportion of all the possible subsets that flow down into each one of the leaves of the trees. By applying this methodology, the computation cost of the algorithm is reduced from exponential to polynomial. With the previous modification, the results are equivalent to evaluating all the possible subsets, $2^N$, in equation (\ref{eq:shapley_values}). Although keeping a record of how many subsets pass through each branch of the tree appears to be simple, it results in subsets of different sizes that are not properly weighted (the weights depend on the number of features present in coalition $S$, which we denote as $\left|S\right|$). The tree-SHAP algorithm corrects the previous issue by extending the subset adding the proportion of ones and zeros (EXTEND method), and reversing it (UNWIND method)~\citep{lundberg2018}. The EXTEND method is applied as the information advances in the tree. The process is reversed through the UNWIND method when the information is split on the same feature twice and it is also applied to undo the extension in a leaf.

The tree-SHAP algorithm can also account for interaction effects. The pairwise interactions are evaluated leading to a matrix of attribution values, where the Shapley interaction index is evaluated in Equation (\ref{eq:shapley_interactions}).

\begin{equation}
    \label{eq:shapley_interactions}
    \phi_{i,j} = \sum_{S\subseteq F\backslash\left\{i,j\right\}}\frac{\left|S\right|!\left(N-\left|S\right|-2\right)!}{2\left(N-1\right)!}\nabla_{i,j}\left(S\right)\text{.}
\end{equation}

\noindent Here, $\phi_{i,j}$ is the interaction SHAP value between features $i$ and $j$, $S$ is a subset of the total possible subsets $F$ that do not contain features $i$ and $j$. The value of the term $\nabla_{i,j}\left(S\right)$ is provided in equation (\ref{eq:shapley_interactions_2}).

\begin{equation}
    \label{eq:shapley_interactions_2}
   \nabla_{i,j}= f\left(S\cup\left\{i,j\right\}\right)-f\left(S\cup i\right)-f\left(S\cup j\right)+f\left(S\right)\text{.}
\end{equation}

The interaction values split symmetrically between both features, with $\phi_{i,j} = \phi_{j,i}$ and the total interaction effect is $\phi_{i,j}+\phi_{j,i}$. Then, the interactions of the importance values captured by the tree and reflected in the SHAP scores can be uncovered by calculating the main effects of a feature for a prediction, as the difference between the SHAP value and its interactions:

\begin{equation}
    \label{eq:shaple_maineffect}
    \phi_{i,i} = \phi_{i}-\sum_{j\neq i}\phi_{i,j}\text{.}
\end{equation}

\subsection{Gradient SHAP}

 For large artificial neural networks with a high number of parameters, the calculation of the kernel SHAP is, computationally speaking, consuming and the tree SHAP cannot be evaluated. For these reasons, the knowledge of the architecture of the model can be exploited to reduce the computational cost of the additive-feature-attribution method. A natural solution is the use of the gradients of the model, {\it i.e.} the gradient-SHAP method. According to \citet{sundararajan2017}, machine-learning practitioners regularly inspect the products of the model coefficients with the feature values to debug the predictions of the model. Computing the gradients of the output with respect to the input is an analog of the previous methodology~\citep{baehrens2010,simonyan2013}. Nevertheless, the gradients break the sensitivity axiom that all the attribution methods should satisfy. This axiom is implicitly included in the previous three axioms and states that if an input and a baseline differ only in one of the features and provide different outputs, then the attribution of the feature must be non-zero. In fact, \citet{sundararajan2017} provided an example to illustrate the previous idea. Let us consider a model defined such as:

\begin{equation}
    \label{eq:examplesensitivity}
    f\left(x\right) = 1-\rm{ReLU}\left(1-x\right)\text{,}
\end{equation}

\noindent where the baseline is $0$, the input $2$ and $\rm{ReLU}$ denotes the rectified-linear-unit activation function. Then the function changes from $0$ to $1$, but $f$ becomes flat at $x=1$, thus, the attribution for $x$ is $0$. 

In order to overcome the previous limitation, \citet{sundararajan2017} proposed the method known as integrated gradients (IG). In this methodology, a straight line path is considered from the reference or baseline $x_{\rm{r}}$ to the input $x$. Then, the gradients along the path are computed, avoiding the problems presented in the previous example. The integrated gradients are computed by accumulating the previous gradients along the path. The integrated gradients are defined in Equation (\ref{eq:integratedgradients}).

\begin{equation}
    \label{eq:integratedgradients}
    {\rm{IG}}_{i} = \left(x_{i}-x_{r_i}\right)\int_{\alpha=0}^1\frac{\partial f\left(x_{r}+\alpha\left(x-x_{r}\right)\right)}{\partial x_{i}}{\rm{d}}\alpha
\end{equation}

The integrated gradients satisfy the completeness axiom, which states that the attributions add the difference between the output of the input and the baseline:

\begin{equation}
    \label{eq:completeness}
    \sum_{i=1}^{n} {\rm{IG}}_{i} = f\left(x\right)-f\left(x_{r}\right)\text{.}
\end{equation}

Fullfilling the completeness attribution implies that the method also satisfies the sensitivity axiom. Therefore, the $\rm{IG}$ method solves the problem of analyzing the gradients of the output. 

Finally, the integral is computed through a summation with sufficiently small intervals from the baseline to the input, as follows:

\begin{equation}
    \label{eq:integratedgradients}
    {\rm{IG}}_{i} \approx \left(x_{i}-x_{r_i}\right)\sum_{k=0}^{m}\frac{\partial f\left(x_{r}+\frac{k}{m}\left(x-x_{r}\right)\right)}{\partial x_{i}} \frac{1}{m}\text{,}
\end{equation}

\noindent where $m$ is the number of steps of the Riemman approximation of the integral. As reported by  \citet{sundararajan2017}, a range from $20$ to $300$ steps is typically enough to approximate the integrals.

The use of integrated gradients implies a reduction of the computational speed of training with back-propagating gradients. Therefore, \citet{erion2021} proposed a new approach to the previous methodology, formulating the integral as an expectation. In addition, this formulation also allows to calculate the importance of each input attribute relative to a batch of baseline inputs, $D$. The expected-gradients methodology accommodates this batch by performing a Monte-Carlo integral, referring $\mathbb{E}$ to the expectation, with samples from the multiple references and interpolation points, $\alpha$:

\begin{equation}
    \label{eq:expectedgradient}
    {\rm{EG}}_{i} = \mathbb{E}_{x_{r}\sim D,\alpha\sim\left\{ 0,1\right\}}\left[\left(x_{i}-x_{r_{i}}\right)\frac{\partial f\left(x_{r}+\alpha\left(x-x_{r}\right)\right)}{\partial x_{i}}\right]\text{.}
\end{equation}

The expected gradients (EG) are used for the calculation of the SHAP values of equation (\ref{eq:additivefeatureattributionmethods}) in the case of the gradient-SHAP algorithm, {\it i.e.} ${\rm{EG}}_{i} = \phi_{i}$.

\subsection{Deep SHAP}

In the case of the gradient SHAP, the expectation of the gradients is calculated by selecting a random set of possible intermediate states of the straight line path between the reference and the input. Deep SHAP approximates the previous calculation by replacing the gradient at each intermediate state with its average values in a single step~\citep{ancona2017}. The SHAP values are calculated assuming that the features are independent and the model is linear. Deep SHAP is equivalent to linearizing the non-linear components of a neural network~\citep{lundberg2017}. This effect is obtained by combining the DeepLIFT (learning important features)~\citep{shrikumar2016} algorithm with the Shapley values.

The DeepLIFT algorithm attributes a value $C_{\Delta x_{i},\Delta o}$ to each input $x_{i}$. This attribution represents the value of an input being set to the original value instead of a reference:

\begin{equation}
    \label{eq:deeplift}
    \sum_{i=1}^{n} C_{\Delta x_{i},\Delta o} = \Delta o\text{,}
\end{equation}

\noindent where $n$ is the number of inputs, $o$ the output, $\Delta o = f\left(x\right)-f\left(x_{r}\right)$, and $\Delta x_{i} = x-x_{r}$. The DeepLIFT method is a modification of the additive-feature-attribution method defined in equation (\ref{eq:additivefeatureattributionmethods}), where $C_{\Delta x_{i},\Delta o}$ replaces $\phi_{i}$ and $f\left(x_{r}\right)$ replaces $\phi_0$. The DeepLIFT algorithm satisfies the local accuracy and the missingness properties. However, it needs to be combined with the Shapley values to satisfy the consistency property. 

The deep-SHAP method calculates the SHAP values by combining the corresponding SHAP values of the smaller components. The algorithm recursively passes the DeepLIFT multipliers to calculate the SHAP values backwards to the network. The SHAP values through every layer are calculated linearly as follows:

\begin{equation}
    \label{eq:linearshap}
    \phi_{i}\left(f_{j},x\right) \approx m_{x_{i},f_{j}}\left(x_{i}-\mathbb{E}\left[x_{i}\right]\right)\text{.}
\end{equation}

\begin{figure}[h]
    \centering
    \includegraphics[width=1\linewidth]{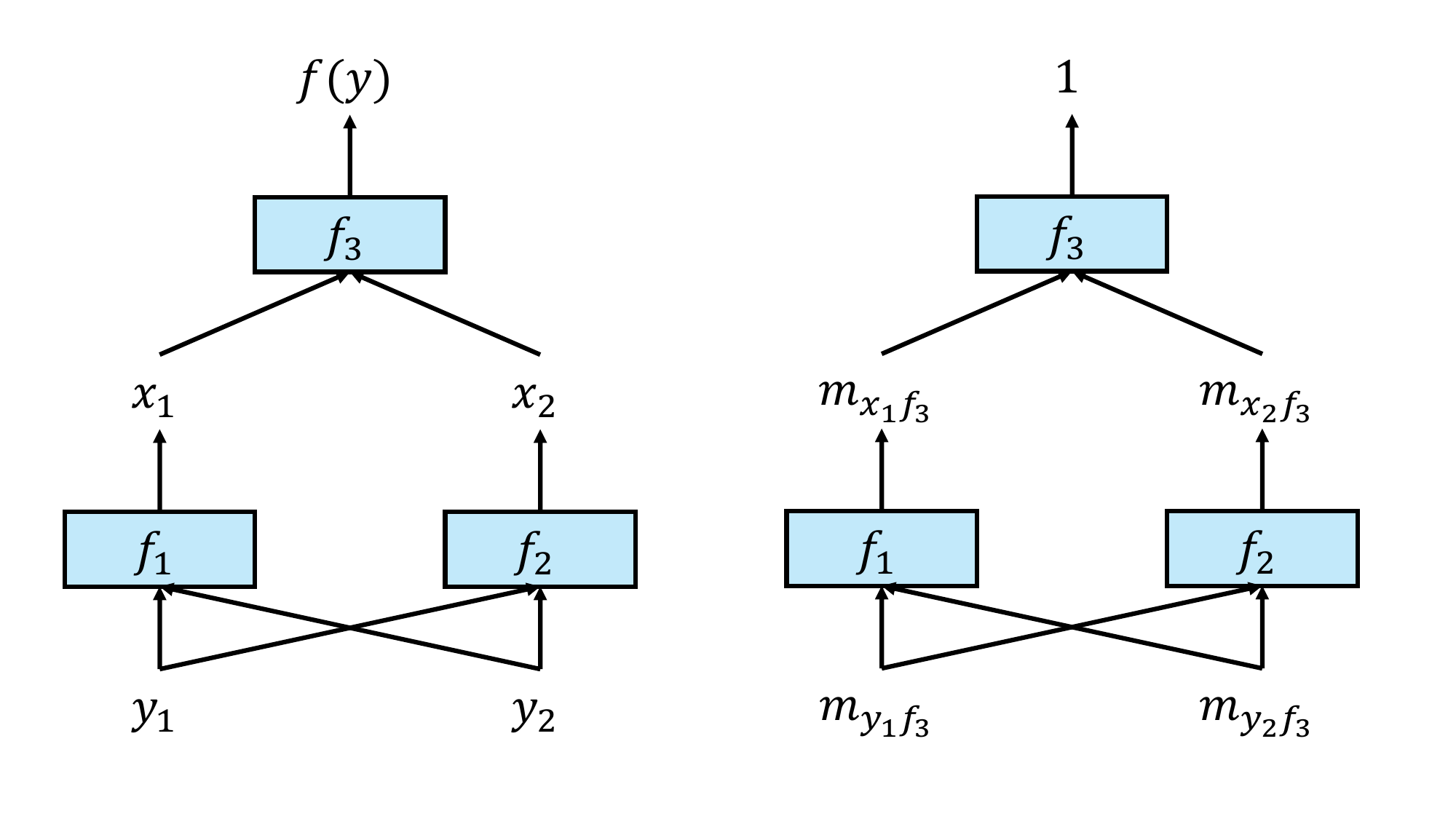}
    \caption{Example model to illustrate the deep-SHAP} methodology. In the figure, the individual component $k$ of the network is represented by $f_k$, the input feature $i$ by $y_i$, its attribution in the component $k$ of the network by $m_{y_i,f_k}$, the $j^{\rm{th}}$ hidden layer value by $x_j$, and its attribution for the component $k$ by $m_{x_j,f_k}$. Figure adapted from \citet{lundberg2017} with permission from the publisher (ACM Digital Library).
    \label{fig:figure17}
\end{figure}

In order to clarify the previous methodology, \citet{lundberg2017} proposed a small example represented in Figure \ref{fig:figure17}.

The SHAP values are calculated recursively according to in the following expressions, where the input features are represented by $y_i$, the hidden features by $x_j$ and the simple components of the network by $f_k$:

\begin{equation}
    \label{eq:deepshap_1}
    m_{x_j,f_3} = \frac{\phi_i\left(f_3,x\right)}{x_j-\mathbb{E}\left[x_j\right]}\text{,}
\end{equation}
\begin{equation}
    \label{eq:deepshap_2}
    \forall_{j\in\left\{1,2\right\}} \; m_{y_i,f_j}=\frac{\phi_i\left(f_j,y\right)}{y_i-\mathbb{E}\left[y_i\right]}\text{,}
\end{equation}

\noindent and then, the chain rule is applied:

\begin{equation}
    \label{eq:deepshap_3}
    m_{y_i,f_3} = \sum_{j=1}^{2}m_{y_i,f_j}m_{x_j,f_3}\text{.}
\end{equation}

\noindent Finally, the SHAP values are calculated using the linear approximation:

\begin{equation}
    \label{eq:deepshap_4}
    \phi_i\left(f_3,y\right) \approx m_{y_i,f_3}\left(y_i - \mathbb{E}\left[y_i\right]\right)\text{.}
\end{equation}

The deep-SHAP method allows a fast approximation of the whole model. The SHAP values of the simple network components, $f_j$, can be calculated analytically if the components are linear, max pooling, or activation functions depending only on one input.

\section{Additive-feature-attribution methods applied to fluid mechanics}

Since the decade of 1990, machine learning (ML) has been applied to modeling fluid dynamics~\citep{faller1997,karunanithi1994}. However, for most of the ML models, their outputs are employed without focusing on the relationships between the inputs and the outputs, which remain unclear. This black-box behavior is one of the main drawbacks of machine learning~\citep{ribeiro2016,vinuesa2021_2}, and thus, adding an extra explainable layer through additive-feature-attribution methods improves the interpretation of the model, showing the impact of the input features. 

Similarly, as occurred for the application of machine learning in fluid mechanics, the explainability of the machine-learning models in the context of fluid dynamics revolves around three main goals. The first goal is using the additive-feature-attribution methods for improving turbulence modeling, mainly Reynolds-averaged Navier--Stokes~\citep{reynolds1895,LAUNDER1974} (RANS) closure equation~\citep{scillitoe2021,mandler2023,bounds2024}. The second is understanding fundamental aspects of fluid mechanics, and more concretely of wall-bounded turbulence~\citep{cremades2024}. Finally, the last category includes particular applications of fluid mechanics such as the evaluation of turbulence in the gliding slope of an aircraft~\citep{khattak2023}, the determination of the pressure coefficient on a building surface~\citep{meddage2022} or the identification of two-phase-flow regimes~\citep{khan2023}.

\subsection*{Additive-feature-attribution methods to improve turbulence modeling}

\begin{figure*}[h]
    \centering
    \includegraphics[width=0.75\textwidth]{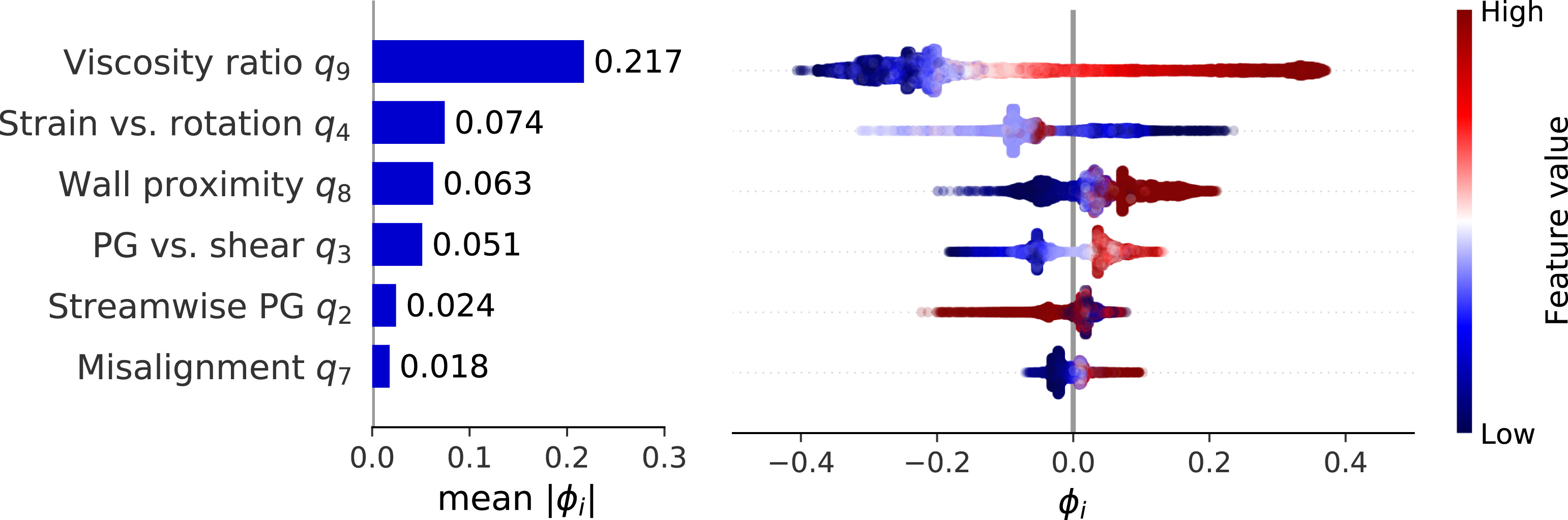}
    \caption{Summary of the SHAP analysis of the input features for predicting the eddy viscosity. The figure shows the mean absolute SHAP values (left) and bee swarm plots of SHAP values (right). Each dot represents a single grid point, when multiple points are located in the same horizontal position, they pile up vertically to show the probability density. The present figure shows the value for the viscosity ratio $q_9$, the strain-rotation ratio $q_4$, the wall proximity $q_8$, the pressure gradient-shear ratio $q_3$, the streamwise pressure gradient normalized with the local kinetic energy and the Reynolds number per unit of length $q_2$ as well as the misalignment of the velocity vector with respect to the streamline velocity. Figure extracted from \citet{he2022} with permission from the publisher (Elsevier).}
    \label{fig:figure1}
\end{figure*}
%$\star$We need to check it$\star$}

RANS models are widely used in industry due to their computational efficiency~\citep{menter2011}. They rely on the Reynolds decomposition and on keeping fluctuations of the velocity in the problem by using the Reynolds stresses. To solve these Reynolds stresses, the eddy viscosity is introduced in the equations to link them with the mean velocity~\citep{boussinesq1903}. 

However, although the traditional RANS models are widely used, they exhibit reduced accuracy in complex flows due to their simplicity~\citep{pop00,slotnick2014}. Therefore, due to the capacity of machine learning models to adapt to complex flow conditions such as adverse pressure gradients, complicated geometries, and separated flows~\citep{vinuesa2022}, many authors have proposed different data-driven methods for turbulence modeling~\citep{duraisamy2019}. However, most of these works do not establish a link between the closure terms predicted by the models and the turbulent input features. To improve the explainability of the models, a recent analysis adds an explainability stage to the research, linking the results of the machine learning model to the physics of the flow~\citep{mcconkey2022,sudharsun2023}.

 On the one hand, the previous works share some similarities to analyze the effect of the turbulent features of the flow to improve turbulent models such as Spalart--Allmaras~\citep{spalart1992} and shear-stress-transport (SST) models~\citep{menter1994}. Firstly, for the inputs, they use the physical properties of low-fidelity flows. For instance, the work of \citet{he2022} employs magnitudes related to the pressure gradient, strain-rotation features, misalignment of the velocity, turbulence legth scales and viscosity ratio between turbulent and laminar values. Then, their mapping between the input features and the corrections of the turbulent models uses a machine-learning model, varying from neural networks~\citep{rosenblatt1958} to random forests~\citep{breiman2001}.

\begin{figure}[h]
    \centering
    \includegraphics[width=0.9\linewidth]{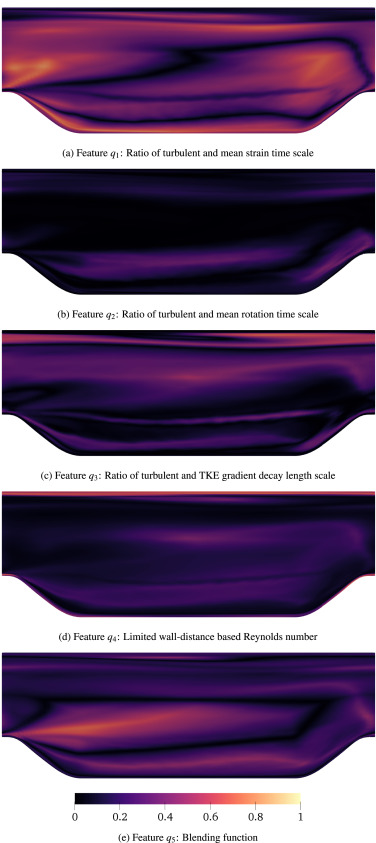}
    \caption{Connection of the SHAP values (importance scores) of each flow magnitude used for the prediction of the turbulence model with the corresponding grid point of the domain. The figure shows the normalized local feature importance scores of the neural-network model based on the deep-SHAP attribution method. Figure extracted from \citet{mandler2023} with permission from the publisher (Elsevier).}
    \label{fig:figure9}
\end{figure}

 Finally, the SHAP values are evaluated, scoring the importance of each physical input parameter to the estimated corrections~\citep{bhushan2023,wu2023}. These importance scores are exemplified in Figure \ref{fig:figure1}, which shows the scores of the viscosity ratio, the strain-rotation ratio, the pressures gradient-shear ratio, the streamwise pressure gradient normalized with the local kinetic energy and Reynolds number per unit length and the misalignment of the velocity vector concerning the streamlines~\citep{he2022}. The previously mentioned figure shows the viscosity ratio to be the most influential flow magnitude followed by the strain rotation ratio. In addition, the figure links the importance score of each grid point (points in the right figure) to the value of the input variable, showing how the SHAP framework reduces nonlinear and complex relationships to a parameter with a simple interpretation. In addition, the SHAP values allow us to detect which input variables are the most important for each region of the domain and whether they present direct or inverse relationships. This idea is exemplified in Figure \ref{fig:figure9}, where the local importance of the five input features is presented as a field in the domain. In the figure, the input feature $q_2$ is the least important feature, while the input features $q_4$ and $q_5$ are the most influential near the upper wall, $q_1$ and $q_5$ are the most influential in the free-shear layer and the recirculation zone.

Other authors such as \citet{fiore2022,mandler2023} exploited the knowledge of the architecture of the model (fully connected neural networks) to evaluate the importance of the turbulent features when modeling the turbulence-closure equations. The former focuses on the prediction of an optimal viscous viscosity for the SST models by using the deep-SHAP~\citep{lundberg2017} algorithm to accelerate the calculation of the importance scores of the turbulent inputs. The latter used the integrated-gradients method~\citep{sundararajan2017} to interpret the relationship between the input variables and their effect on the prediction of the turbulent heat flux for low-Prandtl-number flows.

Additionally, some engineering problems require specific solutions for turbulence modeling. This approach is followed by \citet{wu2023} and \citet{xu2024}. While the former work focuses on the explainability of the turbulence models of a stratified water tank, the second one focuses on the impact on the flow through an axial compressor. Other authors, such as \citet{ayodeji2022}, have applied the SHAP values to the understanding of the closure models for nuclear-power reactors.

\subsection*{Additive-feature-attribution methods to study fundamentals}

\begin{figure}[h]
    \centering
    \includegraphics[width=0.85\linewidth]{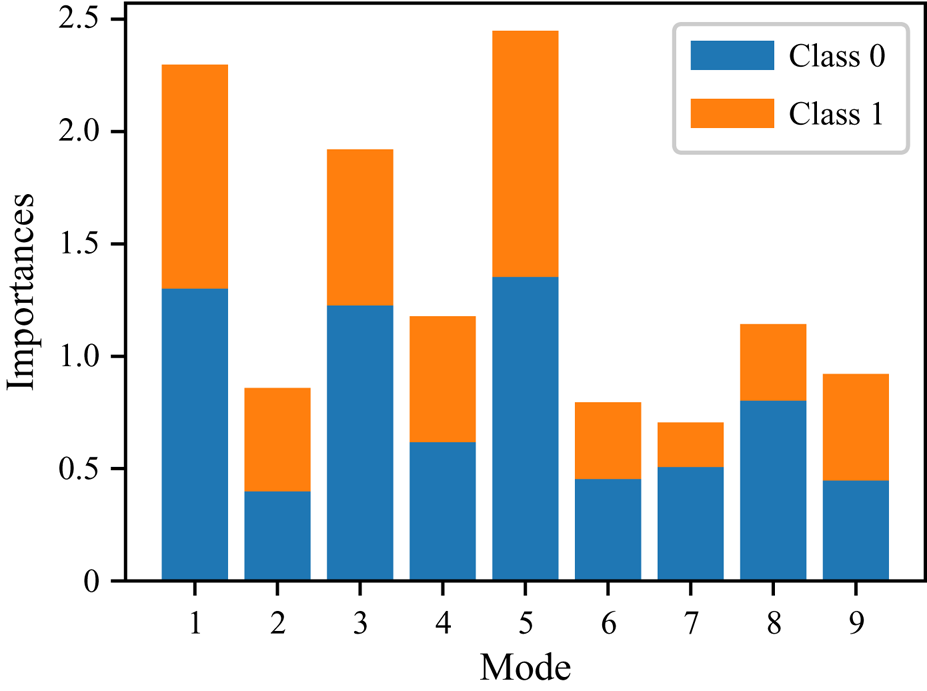}
    \caption{Mean absolute SHAP values for the nine modes of the model in a certain time of simulation. The figure shows the results for both, samples that relaminarize (class 1) and do not relaminarize (class 0). Figure extracted from \citet{lellep2022} with permission from the publisher (Cambridge University Press).}
    \label{fig:figure3}
\end{figure}

\begin{figure*}[h]
    \centering
    \includegraphics[width=1\textwidth]{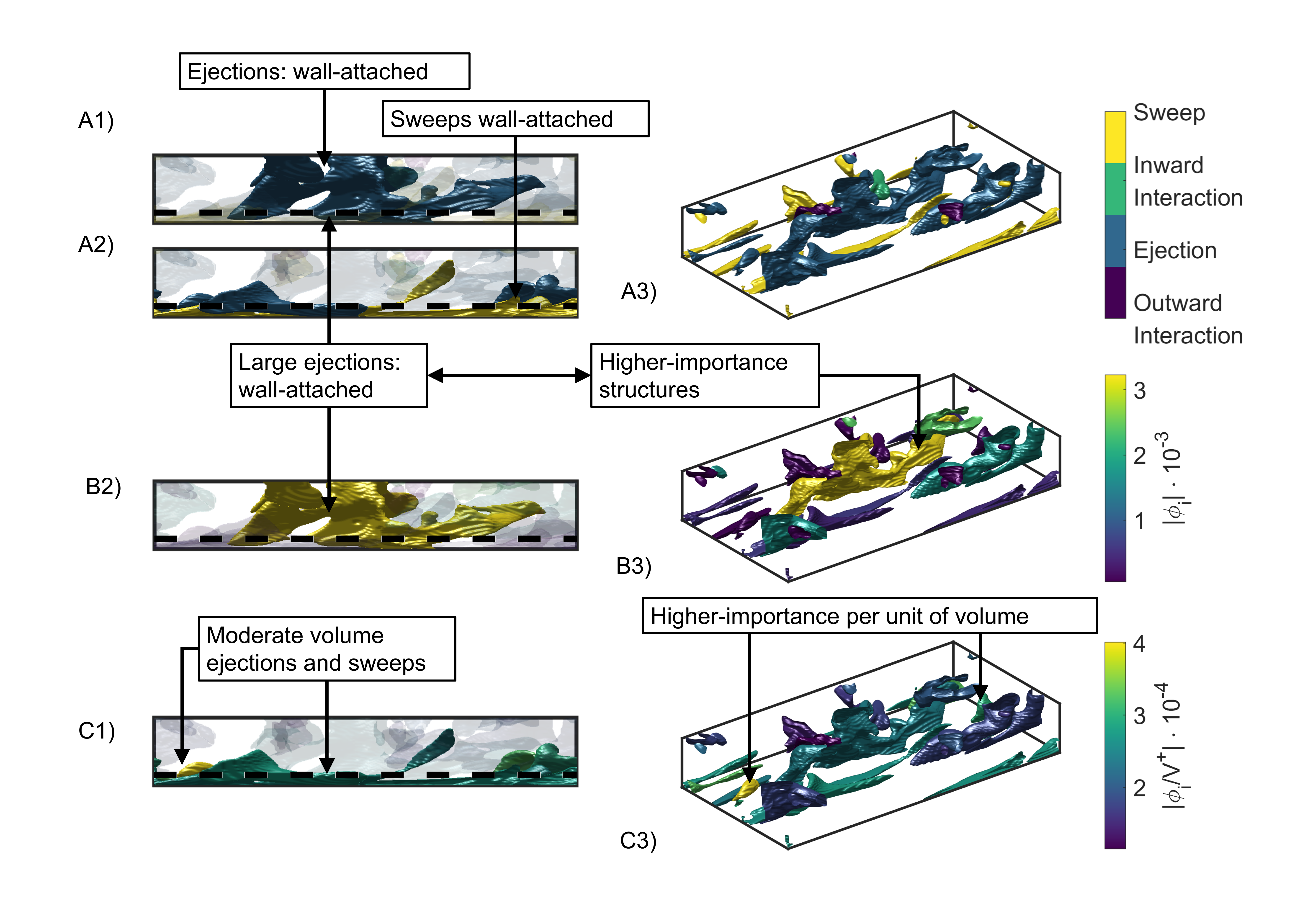}
    \caption{Instantaneous visualization of the intense Reynolds stress structures. This Figure shows (views A) the type of turbulent structure, (views B) the SHAP (Shappley additive explanation) values ($|\phi_i|$) and (views C) the SHAP values divided by the volume ($|\phi_i/V^+|$) of the corresponding structures.  The dashed line marks $y^+=20$, which separates wall-attached and wall-detached structures. Figure extracted from \citet{cremades2024} with permission from the publisher (Springer Nature).}
    \label{fig:figure4}
\end{figure*}

Fluid dynamics are governed by the Navier--Stokes equations, a set of partial differential equations the analytical solution of which has not been found. Due to this limitation and the multi-scale nature of the turbulent phenomena~\citep{kol41a,Cardesa_science}, the computational requirements of integrating numerically these equations are very high~\citep{pirozzoli2004,iwamoto2004,lee15,hoy22}. For these reasons, the capabilities of the additive-feature-attribution methods for detecting the relationships between the input features and the output of a machine-learning model can be used for improving the understanding and explainability of the physics underlying the equations of fluid dynamics, providing new answers and horizons for turbulence analysis. 
An example is the work of \citet{lellep2022} where a nine-mode model is used to simulate a Couette flow with free-slip boundaries~\citep{moehlis2004}. Note that this model has been extensively analyzed in the context of ML research~\citep{srinivasan2019}. The influence of every single mode in the relaminarization of the flow is evaluated by using an XGBoost tree model~\citep{chen2016} analyzed with the previously explained SHAP methodology. The analysis of the additive-feature-attribution methods is exemplified in Figure \ref{fig:figure3}. The figure shows the main modes affecting the relaminarization of the flow. These results evidence that the first mode (laminar profile), the third mode (streamwise vortex), and the fifth mode (specific streak instability) are the most influential in the predictions.

Following the same idea, in the work of \citet{cremades2024} the influence of the intense Reynolds stress structures or Q events~\citep{Lozano2012,Jimenez2018} in a turbulent channel was evaluated. First, the evolution of the flow in the turbulent channel was calculated using a U-net model~\citep{ronneberger2015}. Then, the SHAP values were used to evaluate the impact that each Q event produces on the flow. This impact is calculated using the mean-squared error in the predictions as the output of the model when applying the kernel-SHAP methodology. A visualization of the results is provided in Figure \ref{fig:figure4}. In this figure, the type of Q event of the structures is presented in the group of figures A), their SHAP value in figure B), and the SHAP value per unit of volume in figure C). By applying the SHAP values to the evolution of the turbulent channel flow, the expected results presented by \citet{Lozano2012} were corroborated: wall-attached ejections (structures moving from the wall in the upwind direction) and sweeps (structures moving to the wall in the downstream direction) are the most influential ones. In addition, although the larger structures exhibit a higher SHAP value, when dividing the structures by their volume, different trends were observed. The results of applying the additive-feature-attribution methods to the channel-flow problem revealed that the SHAP values per unit volume are not correlated with the Reynolds stresses per unit volume, evidencing the ability of the SHAP values to detect complex relationships between the inputs and the outputs. These results have been extended by \citet{Hoyas2024} who applied the methodology to other coherent structures (ejections, sweeps, high-velocity streaks, low-velocity streaks, and vortices). In this work, the importance is measured using the impact of the various structures on the prediction of the skin-friction coefficient. 

The difference between the SHAP values and the Reynolds stress indicates that an objective measure of importance should be used for identifying high-impact regions of the turbulent channel. This new concept is explored in the work of \citet{Cremades2024_2}, where the gradient SHAP is used to generate importance fields in the domain of the turbulent channel. Then, the domain is segmented through a percolation analysis, generating structures of high importance, which are then compared with the previously mentioned classical structures studied in the turbulence literature.

\subsection*{Additive-feature-attribution methods for applied problems in fluid dynamics and heat transfer}

Explainable machine learning, and specifically the additi\-ve-feature-attribution methods, provides a clear and interpretable analysis of the impact that a certain input feature has on the output prediction. Although in the previous sections the SHAP values have been applied to the modeling and analysis of turbulent phenomena, there is a wide range of applications directly related to fluid mechanics and heat transfer in which the use of the SHAP values provides insight into phenomena and operation conditions that are complicated to study analytically. These works are related to different fields in which fluid mechanics plays an important role. For instance, the metal industry~\citep{wu2023_2}, yarn production~\citep{koetzsch2024}, flight operations \citep{khattak2024}, renewable energy~\citep{siddiqa2024} or physics \citep{pierzyna2024}. Some of these works have focused on understanding complex physical phenomena, others tried to optimize the design of specific devices or have measured how the deviation in the design parameters affects the performance of the system. In this section, some of the main uses of SHAP values to applied problems in fluid mechanics and heat transfer are presented. The following applications are reviewed next: impact of the atmosphere in human activities, optimization of heat transfer, evaluation of risks and safety improvement, emission reduction and noise mitigation, optimization of rotating machines, and explanation of complex flows.

\subsubsection*{Impact of the atmosphere in human activities}

Regarding the analysis of the atmosphere and its impact on human activities, additive-feature-attribution methods have been applied to different analyses. A particular example of the capabilities of explainable artificial intelligence to understand how atmospheric phenomena impact human activities is related to aviation. The work of \citet{khattak2024_2} focuses on which factors determine the presence of low-level turbulence or those chaotic shifts in headwinds along the airport runway glide path. According to the authors, this phenomenon represents one of the imminent hazards to aviation safety. Low-level turbulence is mainly reported in those airports that are located in regions with rugged mountains. For instance, this phenomenon strongly affects Hong Kong International Airport, being present in one of every 2000 flights. Low-level turbulence can alter the aerodynamic forces that an aircraft needs during landing, leading to safety issues. In the work of \citet{khan2023} the turbulence intensity of the atmosphere in the glide slope of the runway is predicted as a function of a set of input parameters. The terrain effects, the wind direction, the runway orientation, and the distance from the runway are employed for that purpose, using a random-forest-based method to calculate the predictions. The additive-feature-attribution method, or SHAP interpretation, is applied to the previous parameters. The terrain was determined to be the most important feature affecting the turbulence intensity, being more influential than the distance from the runway and the wind direction.

\begin{figure*}[h]
    \centering
    \includegraphics[width=0.8\textwidth]{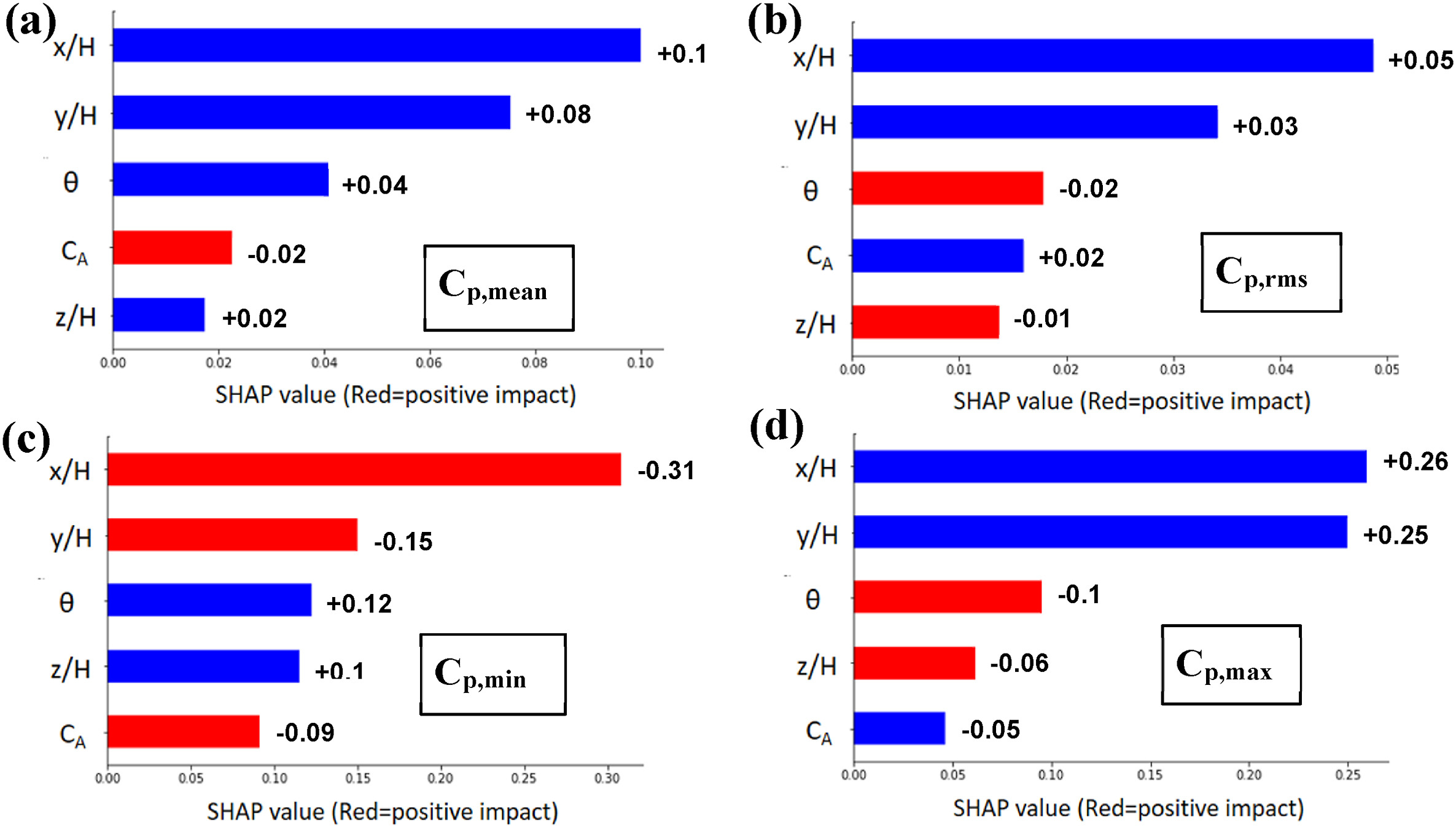}
    \caption{Mean SHAP values of the geometrical inputs of the building for the prediction of the mean, root-mean-squared, minimum and maximum values of the pressure coefficient. The input features are the horizontal, $x/H$ and $y/H$, and the vertical, $z/H$, position of the probes, where $H$ is the height of the building, the wind incidence angle $\theta$ and the area density $C_{\rm{A}}$ (which represents how densely the buildings are arranged). Figure extracted from \citet{meddage2022} with permission from the publisher (Elsevier).}
    \label{fig:figure13}
\end{figure*}

% Civil engineering
In addition to the effects that the wind has on aviation, other industries are affected by the turbulence and instabilities of the atmosphere. Wind loads affect many industries, being civil engineering a clear example of how aerodynamic forces condition the design of structures. The knowledge of the aerodynamic coefficients of the wind loads over the building surfaces is required to size the structure~\citep{tanaka2012} and to estimate its natural ventilation rate~\citep{charisi2021}. Buildings are bluff bodies, therefore, the calculation requires computational fluid dynamics or experimental measurement of the aerodynamic coefficients leading to expensive, time-consuming tests. For the latter specific facilities are also required, a fact that is not always aligned with the commercial interests of the contractors. Tremendous progress in this direction has been reached by applying artificial intelligence to the prediction of aerodynamic loads on buildings~\citep{chen2002,kwatra2002,Vishwasrao2024,nilsson2024}. The effects that the atmosphere has on the buildings are complex and the machine-learning algorithms are difficult to interpret. This is the main reason some authors such as ~\citet{meddage2022} used the additive-feature-attribution methods to understand the relation of the design parameters of the building with the mean, fluctuating, minimum, and maximum pressure coefficients over the external surfaces of a building. This building is located inside an urban environment, creating complex flows and patterns. Therefore, nonlinear relationships are generated between the inputs and the outputs during the training process of the model. For this reason, the use of the SHAP values is essential for simplifying this relationship and linking the geometrical parameters with the expected pressure coefficient. The exemplification of the evaluation of the importance of the input features in the prediction of the pressure coefficient can be observed in Figure \ref{fig:figure13}, where the horizontal coordinates of the probe, $x/H$ and $y/H$ (see figure caption for more details), are demonstrated to be the most influential parameters. A similar analysis was performed by \citet{yan2024}, who analyzed the pressure coefficients in a building facade when it interacts with other buildings. The authors indicated the importance of knowing the wind pressure distributions to avoid safety problems. In that work, the importance of the different geometrical features for the pressure-coefficient prediction was analyzed, demonstrating that the selection of the facade is the most important feature, followed by its position and the height ratio.

\begin{figure*}[h]
    \centering
    \includegraphics[width=1\textwidth]{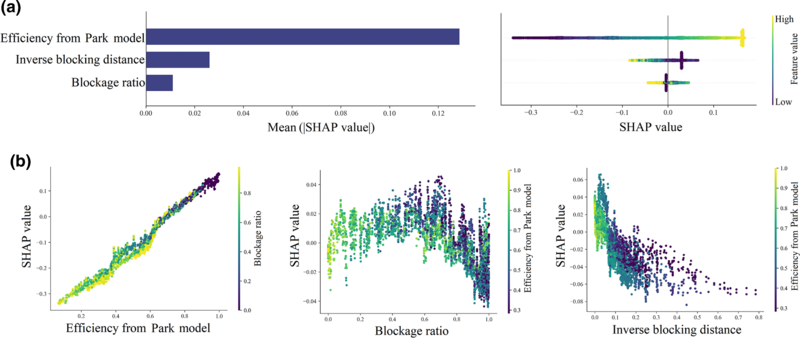}
    \caption{SHAP contribution of the efficiency of the turbine, the inverse blocking distance and the blockage ratio. The bar plot (figure a) shows the mean SHAP of each feature, while the scatter plots (figure b) show the relationship between the input features and the SHAP values. Figure extracted from \citet{zehtabiyan2023} with permission from the publisher (American Institute of Physics).}
    \label{fig:figure10}
\end{figure*}

% Energy engineering
Wind farms are other applications in which the influence of the atmosphere and the wind is crucial for the proper design of the system. The computational simulation of the wind farms is complex and requires a very high computational cost. Due to these requirements, for most industrial purposes simplified models, such as wake models, are widely used~\citep{archer2018}. These methodologies require the superposition of the calculation of each single-turbine wake, applying simplistic physical assumptions which can lead to large errors~\citep{lissaman1979}. To solve this problem, efforts have been made to develop lightweight models based on ML methods~\citep{zehtabiyan2022}. However, the ML methods applied so far in this context rely on the optimization of the solution in terms of the training data, which does not provide flexibility for extrapolation. The work of \citet{zehtabiyan2023} attempts to overcome the previous limitation by improving the explainability of these machine-learning methods. The SHAP values are used to connect the prediction of the efficiency of the turbines with the input features. For these input features, both geometric information about how the turbines are spaced concerning the wind in the different directions (blockage ratio and inverse blocking distance)~\citep{niayifar2016} and turbine-level efficiency are required. The turbine level of efficiency is calculated using the Park model~\citep{katic1987}, based on the \citet{jensen1983} analytical wake model, assuming a top-hat profile for the velocity deficit in the wake of a turbine, as well as the NP model~\citep{niayifar2016}, which extends the Gaussian wake model derived by \citet{bastankhah2014} from the mass and momentum conservation. The analysis of the SHAP values evidences the high importance of the efficiency of each single turbine, as can be observed in Figure \ref{fig:figure10}, where a good correlation between the SHAP values and the efficiency of the turbines is observed. The power generated by a single wind turbine before installation was analyzed by \citet{cakiroglu2024}. The model took the environmental conditions of the atmosphere, such as the velocity, pressure, temperature, or density, and generated the output, {\it i.e.}, the generated power. The SHAP values evidenced the high influence of the wind velocity in the solution, corroborating the analytical models widely used in wind-power production~\citep{manwell2010}.

The atmosphere is a complex environment, where very large scales related to the geographic position of the analysis interact with smaller scales in the walls of the objects located on the surface of the Earth. In addition, these smaller scales condition the friction and performance of the systems while the large scales condition the direction of the freestream conditions. Therefore, machine learning is a useful tool for the estimation of the conditions in which human activity is developed for wind farms, and urban air flows, among others. The application of the SHAP techniques to the atmospheric effect on the human-made environments improves the interpretation of the nonlinear relationships detected by the models, and thus, allows a better operational optimization of industrial activity~\citep{gijon2024}.

\subsubsection*{Optimization of heat-transfer devices}

\begin{figure}[h]
    \centering
    \includegraphics[width=0.95\linewidth]{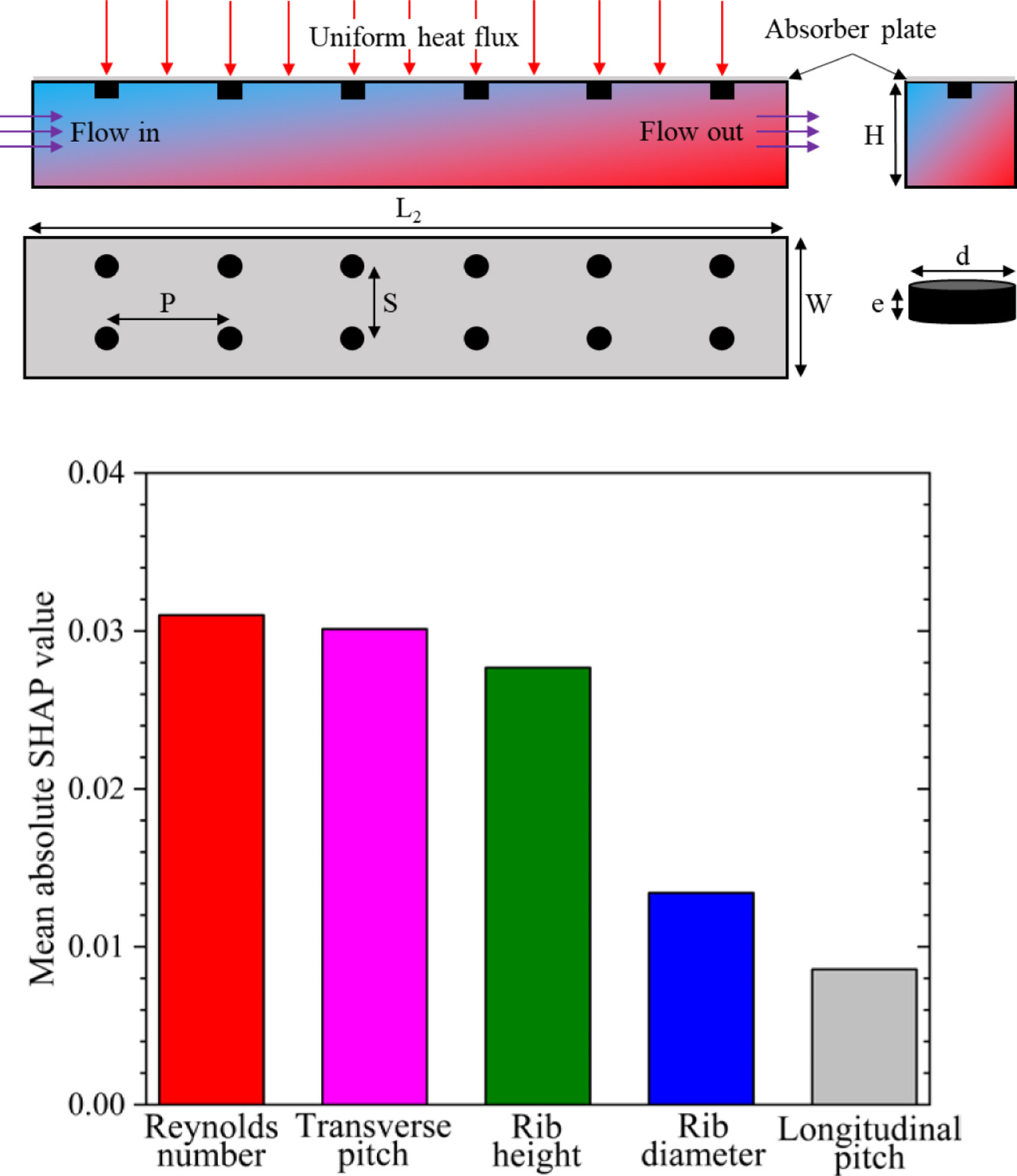}%,trim={0.5cm 0 7cm 0cm},clip
    \caption{Mean absolute SHAP values for the design parameters of the solar air heater: Reynolds number, transverse pitch, $S$, longitudinal pitch, $P$, rib height, $e$, and rib diameter, $d$. Figure extracted and adapted from \citet{singh2022} with permission from the publisher (Elsevier).}
    \label{fig:figure11}
\end{figure}

% heat transfer
Another common application for explainable artificial intelligence in fluid mechanics is heat transfer. The environmental challenges, the reduction of energy consumption, and the use of fossil fuels require the optimization of a wide range of energy-related systems~\citep{liu2013,chen2017,yari2019}. There is a wide range of works focusing on the thermal performance of different devices. For instance, \citet{singh2022} and \citet{suman2024} used the SHAP values to improve the thermal efficiency of solar air heaters. These devices convert solar radiation into useful energy. However, although solar energy is freely available, the performance of these systems is limited by their poor air conductivity and the thermal boundary layer of the absorber plate. These systems are widely used due to their simple design and easy of manufacture. Therefore, to increase the thermal efficiency with minimum changes, artificial roughness is added to the plate. By adding this artificial roughness, the induced turbulence increases the heat-transfer coefficient in the interface of the absorber plate. For including this artificial roughness, transverse ribs are frequently used~\citep{singh2018}, although vertical ribs with different shapes have also been employed~\citep{promvonge2008,manjunath2017}. The geometrical features of the optimum vertical cylindrical ribs were investigated by \citet{singh2022} while \citet{suman2024} analyzed circular ribs. In both cases, the impact of the Reynolds number and the geometrical parameters of the ribs were studied using additive-feature-attribution methods. The SHAP values are calculated to rank the impact of the previous features on the thermodynamic-performance factor, as can be observed in Figure \ref{fig:figure11}.
  
\begin{figure*}[h]
    \centering
    \includegraphics[width=0.85\textwidth]{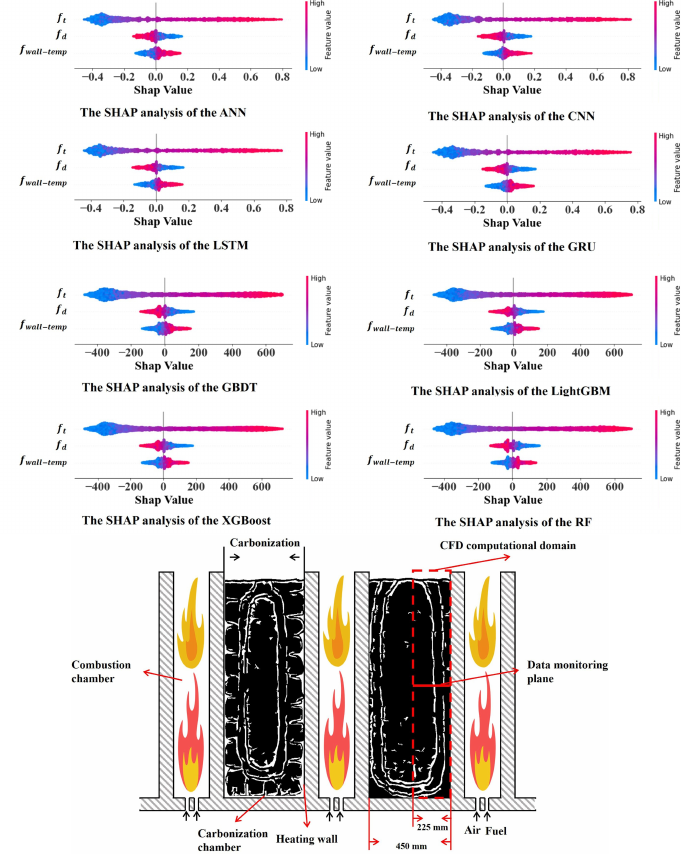}
    \caption{SHAP contribution of the input features in the cokemaking process. The top figure shows the three most influential parameters (coking progress time $f_t$, bulk density $f_d$, and heating wall temperature $f_{\rm{wall-temp}}$) of the eight different models trained by \citet{zhao2024}. For the analysis of the complete set of parameters, the reader is referred to the original paper. The bottom figure shows a schematic representation of the cokemaking process. Figure extracted and edited from \citet{zhao2024} with permission from the publisher (Elsevier).}
    \label{fig:figure12}
\end{figure*}

As previously discussed, an efficient management of the heat transfer during the coking process is essential to reduce energy consumption and to increase the quality of the coke~\citep{tiwari2014,hilding2005}. Cokemaking processes are complex thermochemical conversions in which minor deviations can lead to instabilities~\citep{hilding2005,lee2019}. Due to the difficulty of measuring the temperature in the coke oven, the digital-twin technique is pivotal for the transformation of this industry. The most common method for modeling cokemaking is computational fluid dynamics~\citep{polesek2015,slupik2015}. However, the required numerical simulations are computationally intensive and time-consuming~\citep{marcato2021}. To solve the previous problems, several ML methods have been introduced~\citep{qiu2024}. As stated before, the cokemaking process is complex, and, therefore, to understand the relationships detected by the model, the additive-feature-attribution methods were applied to the model. In fact, \citet{zhao2024} employed the SHAP values to connect the coal properties and the operation conditions with the temperature profile of the coal/coke bed. In this work, the SHAP values were compared across different models. The various models exhibit consistency in the results. As can be observed in Figure \ref{fig:figure12}, the most relevant features are independent of the architecture used for the predictions. However, for the less important features, some disparity can be observed. This fact evidences the independence of the SHAP values from the model. Additive-feature-attribution methods are model-agnostic. Therefore, they do not depend on the structure of the model. Additive-feature-attribution methods can be applied to any type of model since their usage only depends on the relationships detected between the inputs and the outputs. The results by \citet{zhao2024} showed that time is the most important feature for the temperature predictions, being the bulk density, the heating wall temperature, and moisture content also high-impact features.

The fast development of semiconductors has led to more powerful components, and thus, to higher heat production. If heat is not dissipated in time, then, the performance of the electronic components drops drastically~\citep{fan2020}. To evacuate heat in high-power devices, gravity heat pipes are widely used due to their high thermal conductivity, space adaptability, and simple structure. The gravity heat pipes evacuate heat by circulating a coolant fluid from an evaporation section, in which the coolant is boiled to a condensation section, where the energy is released by condensing the vapor~\citep{nikolaenko2021}. These heat pipes were embedded into the radiator structures~\citep{ladekar2023}, enhancing their heat-transfer performance by perforating finned heat sinks~\citep{ibrahim2019}. The optimization of the parameters of the finned heat sinks was analyzed by \citet{wang2023} by means of additive-feature-attribution methods. In this work, the heat source temperature was decreased by 9\% and the heat-sink mass by 18\% compared to an optimal solution calculated in an orthogonal test. \citet{sikirica2023} explored the use of SHAP values for the optimization of microchannel heat sinks~\citep{tuckerman1981}, which evacuate the heat produced by electronic devices by circulating a coolant~\citep{phillips1988}. As stated before, the optimization of these channels is crucial for the computation-power growth of electronic chips, as heat dissipation becomes more challenging for more powerful devices, as it is expected to become a performance-limiting factor. \citet{sikirica2023} proposed a methodology based on machine learning to predict the thermal resistance and pumping power of the microchannel heat sinks. Then, a multi-objective optimization algorithm is used and the most influential variables are obtained using the SHAP values. This SHAP-values analysis evidenced that the width of the channel and the number of secondary channels are the most influential variables for the thermal resistance.

The climate emergency and the necessity of improving the efficiency of industrial systems have required new methodologies for estimating more accurately complex phenomena such as the convective heat exchange on the surface of a solid. Many works have focused on using machine learning to detect the relationships between the design parameters and the heat flow. To understand the influence of the design variables and to optimize the systems, explainable artificial intelligence has been applied to the models.

\subsubsection*{Evaluation of risks and safety improvement}

Although, as previously explained, one of the main applications of fluid mechanics in electric systems is thermal dissipation, which is essential in the device performance, for high-power systems, any possible damage could result in catastrophic consequences. Switchgears are widely used for the control, protection, and isolation of electrical equipment. The architecture of the switchgear uses a gas for isolating the electrical components. An electric arc is formed in the fluid when the electrical current is interrupted. Although a fault in the electric arc is uncommon, it can cause substantial damage including fatal consequences~\citep{kumpulainen2016}. To improve safety, \citet{matin2024} analyzed the parameters influencing the maximum pressure generated during the internal arc in medium-voltage switchgear. A machine-learning model was trained using numerical simulations of the internal arc phenomena. Then, the importance of the manufacturing parameters, such as the duct width and height, and the operational parameters, initial pressure, and temperature, is calculated using the SHAP values. The analysis demonstrated that for the sake of safety, the switchgear manufacturers should increase the area of the ducts, separating them and changing their position instead of increasing their height.

% Safety
A proper understanding of the contribution of the design parameters of the engineering systems is essential for avoiding risks and improving their safety. Another common safety problem in industrial applications is accidental explosions. This safety issue was addressed by \citet{hu2024} for hydrogen-air mixtures. The interest in these mixtures has increased due to the environmental challenge and the requirements to reduce carbon emissions. Hydrogen has a higher energy density than batteries, making it more suitable for new propulsion systems, energy storage, power generation as well as domestic and commercial usages~\citep{hu2023}. However, hydrogen has a high risk of deflagration and detonation, which can cause catastrophic consequences~\citep{park2019}. Hydrogen requires a very low ignition energy in air, 4\% of methane. Therefore, leakage of high-pressure hydrogen can produce self-ignition~\citep{yang2021}. \citet{hu2024} evaluated the features producing accidental explosion loads from hydrogen-air mixtures in vented silos. In the manuscript, the impact of the vent and silo geometrical parameters and the pre-set pressure of the system is evaluated in the prediction of the silo overpressure, the rising rate, and the impulse. The results of the SHAP values confirmed the empirical experience, namely that larger silo diameters imply more development space for a spherical flame, which is accelerated, leading to an increase in the peak overpressure and rising rate. This effect is opposed to a larger silo length and vent size, which discharge gas, reducing the overpressure and the rising rate. Then, the impulse is influenced by the quantity of hydrogen in the combustion. Thus, larger silos have a higher impulse, which, on the other hand, is reduced by larger vents. Finally, the higher the pre-set pressure, the more gas is accumulated in a short period increasing the blast loading.

SHAP values have also been applied to improve the safety in hospitals isolating airborne-infected patients. During the coronavirus pandemic, the medical facilities such as the emergency rooms, played a vital role in the protection against the disease infections~\citep{hobbs2021}. To protect against airborne infections, partitions, and fan-filter units are provided in each bed within emergency rooms~\citep{mousavi2020}. The work of \citet{lee2024} used data-driven models to identify the most important parameters for the design of patient-isolation units. The machine-learning models were trained using the data from numerical simulations. The models evaluated the ventilation performance using input feature variables such as the environmental ventilation system location, the distance between the ventilation system and the patient, or the distance between the areas and the heating, ventilation, and air-conditioning.

The SHAP values have been demonstrated to be a crucial tool in the detection of possible risks in workspaces. They can be used in combination with fluid-flow analysis to detect the most dangerous conditions for the operation of industrial and healthcare facilities. Proper identification of the high-risk parameters results in increased security, which can potentially save lives.

% Continuar aquí
\subsubsection*{Application to pollutant generation and dispersion}

The presence of pollutants and the increase of emissions in human activity present a risk to the population's health. SHAP values have also been applied for detecting the conditions that reduce the presence of pollutants. A proper analysis of these values provides the required information for decision-making and for developing new regulations that can fight environmental challenges and prevent pollution-derived health problems. Human activity has increased dramatically the generation of pollutants worldwide in recent years. An example of these pollutants is nitrogen oxides, $\text{NO}_\text{X}$. In the case of \citet{ye2022}, the generation of $\text{NO}_\text{X}$ in a coal-fired power plant is analyzed using additive-feature-attribution methods. The prediction of the generation of nitrogen oxides is a challenge due to the complex turbulence, chemical reactions, and heat transfer. The production of $\text{NO}_\text{X}$ directly depends on the operation parameters, such as the coal supply, the primary and secondary airflow, or the burnout airflow~\citep{kim2018}. This work used simulated and operational data to train a model for the prediction of the $\text{NO}_\text{X}$. Finally, this prediction is explained by applying the SHAP values to the input features, such as fuel, auxiliary air, or overfire air. The prediction of $\text{NO}_\text{X}$ emission was also analyzed by \citet{ding2023}. In this case, the pollutant emission was studied in the case of a waste incineration power plant. A model was trained using real data from the power plant, then the SHAP values were used for evaluating the $\text{NO}_\text{X}$ emissions as a function of the input features of the model: ammonia flow, $\text{CO}$ concentration in the exhaust smoke, furnace temperature, {\it etc.}. The results suggested that more ammonia ejection can effectively reduce the $\text{NO}_\text{X}$ emission.

On the other hand, other analyses focus on the air pollutant dispersion. An example of explainable artificial intelligence applied to pollutant dispersion was presented by \citet{bai2023}. The work focused on the dispersion on a street canyon, a U-shaped spaced in a street flanked by buildings~\citep{nicholson1975}. The work linked the relationships between the ratio of the building height, the street width, and the building length with the pollutant concentration variation in the street canyons. The results of the SHAP value analysis evidence that the height-width ratio, the placement of the sensor in the along-canyon position, and the temperature are the most influential features. In addition, while the high height-width ratio and the low temperature have a positive contribution to the $\text{CO}_2$ concentration, the position of the sensor impact trend is not proportional. The effect of the environmental and atmospheric variables is also analyzed for the concentration of pollutants. \citet{wang2023_2} studied the  concentration of $\text{PM}_\text{2.5}$ and $\text{O}_\text{3}$ in the Fenwei Plain. In this work, the pollutant data from the National Environmental Monitoring Center of China and the meteorological data from the National Aeronautics and Space Administration were used for the prediction of the previous pollutants. The results evidenced that the particles $\text{PM}_\text{10}$, the $\text{CO}$ and the temperature are the most influential parameter for the prediction of the $\text{PM}_\text{2.5}$, while the net surface solar radiation, the $\text{NO}_\text{2}$ and the $\text{CO}$ affect more the prediction of the $\text{O}_\text{3}$.

\subsubsection*{Noise prediction and reduction}

As previously discussed, the reduction of emissions is crucial for the environmental challenges that humanity needs to face. In addition to pollutants, noise generation must be reduced to achieve environmentally sustainable development~\citep{li2013,mohamed2016}. Aerodynamic noise is one of the main sources of noise in aviation~\citep{lockard2004,zhao2018}, ground transportation~\citep{aabom2014,papoutsis2015} and space sector~\citep{escarti2024}. For these reasons, the aerodynamic self-noise generated by an airfoil was analyzed, using SHAP by \citet{yadam2020}. This work focused on the explainability of the airfoil parameters in predicting sound pressure levels. For this purpose, a machine-learning model was trained using NASA's open-source data set of aerodynamic and acoustic tests conducted in an anechoic wind tunnel~\citep{brooks1989}. The model predicted the sound pressure levels using as input parameters the frequency, the angle of attack, the chord, the free stream velocity, the suction side displacement thickness, and the Reynolds number. The interpretation of the influence of the input features in the output is conducted by calculating the SHAP values of the inputs. The SHAP values evidenced the higher contribution of the frequency and the suction side displacement to the sound pressure levels. The lower values of these features are related to the positive contribution of the sound pressure level, while for less influential features, such as the freestream.

\begin{figure}[h]
    \centering
    \includegraphics[width=1\linewidth,trim={0cm 0 0cm 0cm},clip]{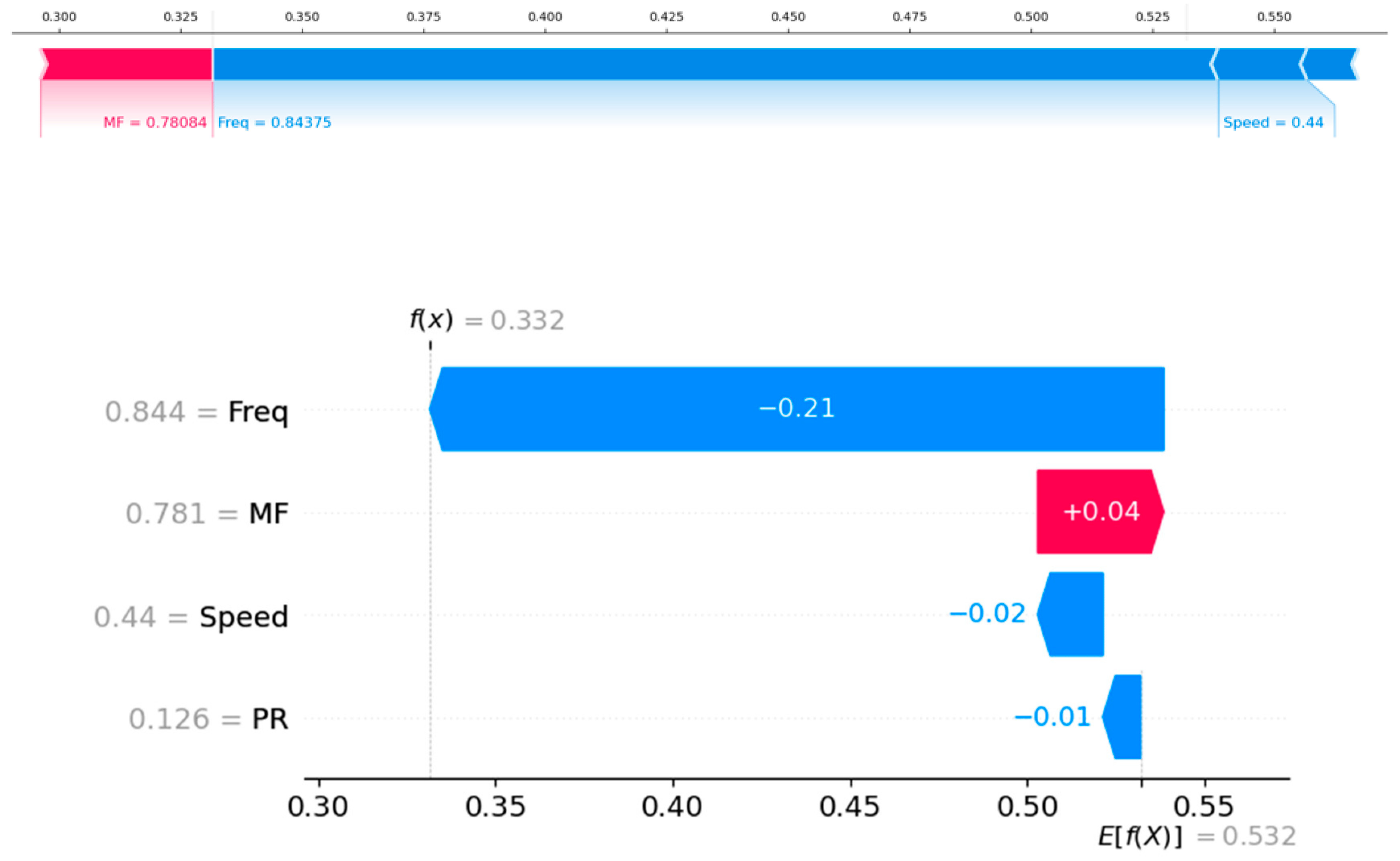}
    \caption{Interpretation of the SHAP values of the different input features for the prediction of the sound pressure level of a turbocharger: Freq (frequency of the noise), MF (mass flow), Speed (speed of rotation of the compressor) and PR (pressure ratio in the compressor). Figure extracted from \citet{huang2023} with permission from the publisher (Multidisciplinary Digital Publishing Institute).}
    \label{fig:figure14}
\end{figure}

The reduction of gas emissions in ground transportation requires increasing the specific power of the engines. This improvement is commonly addressed by using turbochargers~\citep{bao2022,broatch2018}. To understand how the operation parameters affect the noise generation, \citet{huang2023} applied the additive-feature-attribution methods to the turbocharger noise emission problem. A machine learning model was trained, using experimental data, to predict the sound pressure level from the frequency, the speed of the rotor, the mass flow, and the pressure ratio. The results evidenced that the frequency is the most influential parameter for the sound pressure level. Its influence is negative, therefore, low frequencies produce a higher sound pressure level. The interpretation of the SHAP values for the calculation of the sound pressure levels of the turbocharger can be observed in Figure \ref{fig:figure14}.

Therefore, current environmental challenges require noise reduction of the various means of transportation and their subsystems. The noise generation is dependent on the operation conditions. Large databases on the performance of the systems can be used for training models which can be analyzed with additive-feature-attribution methods. The application of these methods to the noise-generation problem provides deeper insights into the reduction of noise emissions.

\subsubsection*{Application to the optimization and understanding of the uncertainty effect of rotating machines}

Rotating machines are crucial for a wide number of propulsive~\citep{gallar2011}, energetic~\citep{huda2014}, and industrial~\citep{liu2020} systems. On the one hand, regarding the optimization of the rotating machines, \citet{fu2024} proposed an optimization methodology using explainable artificial intelligence. In this work, the geometrical parameters of the rotor cage of a straw micro-crusher classification device were studied for increasing its classification performance, which is crucial for achieving the powder quality required for biomass resources~\citep{shapiro2005}. The installed angle length and number of blades are used for predicting the cut size of the particles and the classifying sharpness index. The results of the SHAP analysis evidenced that the installed angle of the blades is the most influential parameter, followed by their length for both cases.

\begin{figure}[h]
    \centering
    \includegraphics[width=1\linewidth,trim={0cm 0 0cm 0cm},clip]{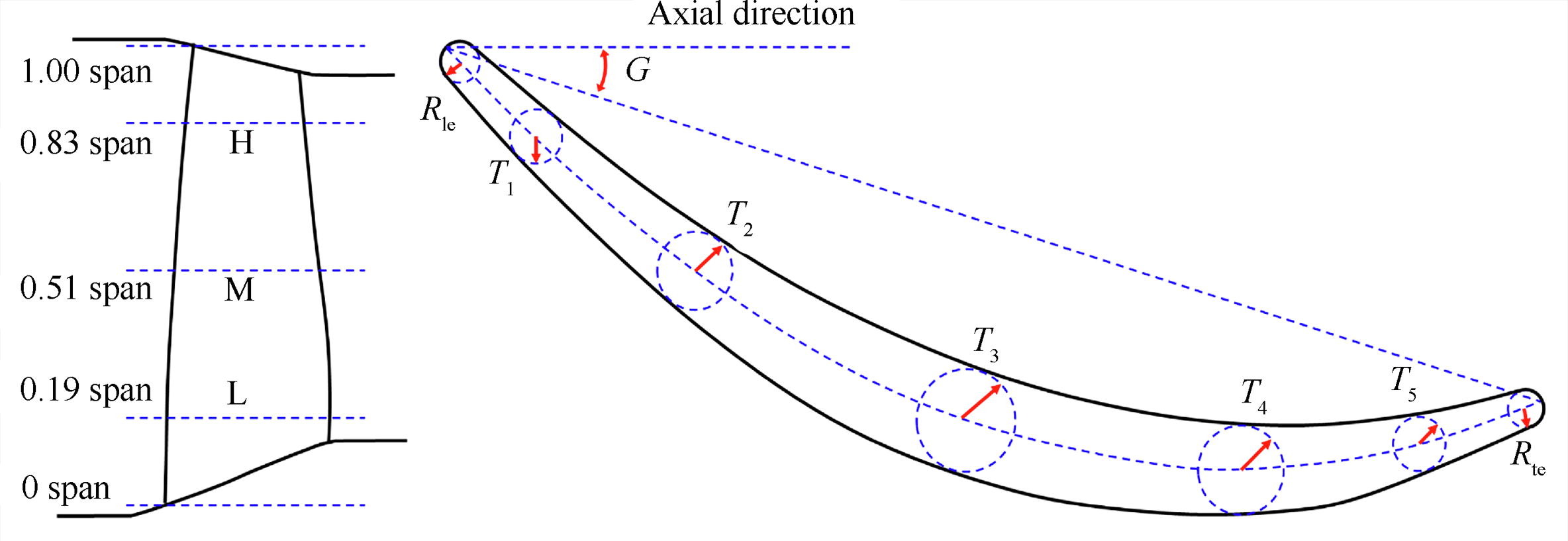}
    \caption{Parameterization of the compressor blade used for generating the database used in the explainability analysis. The angle between the chord and the axial direction $G$, the radius of curvature $R$ of the leading $le$ and trailing $te$ edge and the thickness along the chord $T_i$ are analyzed for three sections $L$, $M$ and $H$ along the blade span. Figure extracted from \citet{junying2023} with permission from the publisher (Chinese Society of Aeronautics and Astronautics).}
    \label{fig:figure15}
\end{figure}

On the other hand, some authors such as \citet{wang2022_2} or \citet{junying2023} identified the importance of understanding the effect that the geometrical uncertainties of a compressor have on the operation conditions. The compressors are a core component in the aero-engines and the gas-turbine stations~\citep{rolls2015}. The performance of the compressor conditions the results of the complete system. Manufacturing errors and degradation might produce deviations in their geometrical characteristics. Thus, finding the critical geometric variables for the performance of the component is of high relevance~\citep{wang2020}. The work of \citet{junying2023} focused on finding the differences in the influence of the manufacturing uncertainties of the geometric variables of a rotor blade under a range of working conditions. The data was generated using computational fluid dynamics to create a database for evaluating the compressor performance as a function of the geometric parameters of a  blade. The parameters are presented in Figure \ref{fig:figure15}, where the angle formed by the chord of the airfoil and the axial direction is denoted by $G$, the radius of curvature of the leading edge by $R_{le}$, and the radius of the trailing edge by $R_{te}$, and the thickness along the chord by $T_i$. 

Then, the geometric variables are extracted from 100 manufactured blades and statistically analyzed. A design of experiments is used for determining the required simulations used to create, assuming Gaussian distributions, the database employed for training the model (see the work by \citet{morita2022} for more information on ML for parametrized geometries). The SHAP values of the machine-learning model are used to extract the influence of the uncertainty variables under six different working conditions. The results showed a high importance of the leading-edge radius for most of the conditions. A different trend is observed for choke conditions in subsonic flows, which are more critically influenced by the thickness and curvature of the airfoils. A similar procedure was presented by \citet{wang2022_2} for a three-stage compressor.

\begin{figure*}[h]
    \centering
    \includegraphics[width=0.94\textwidth]{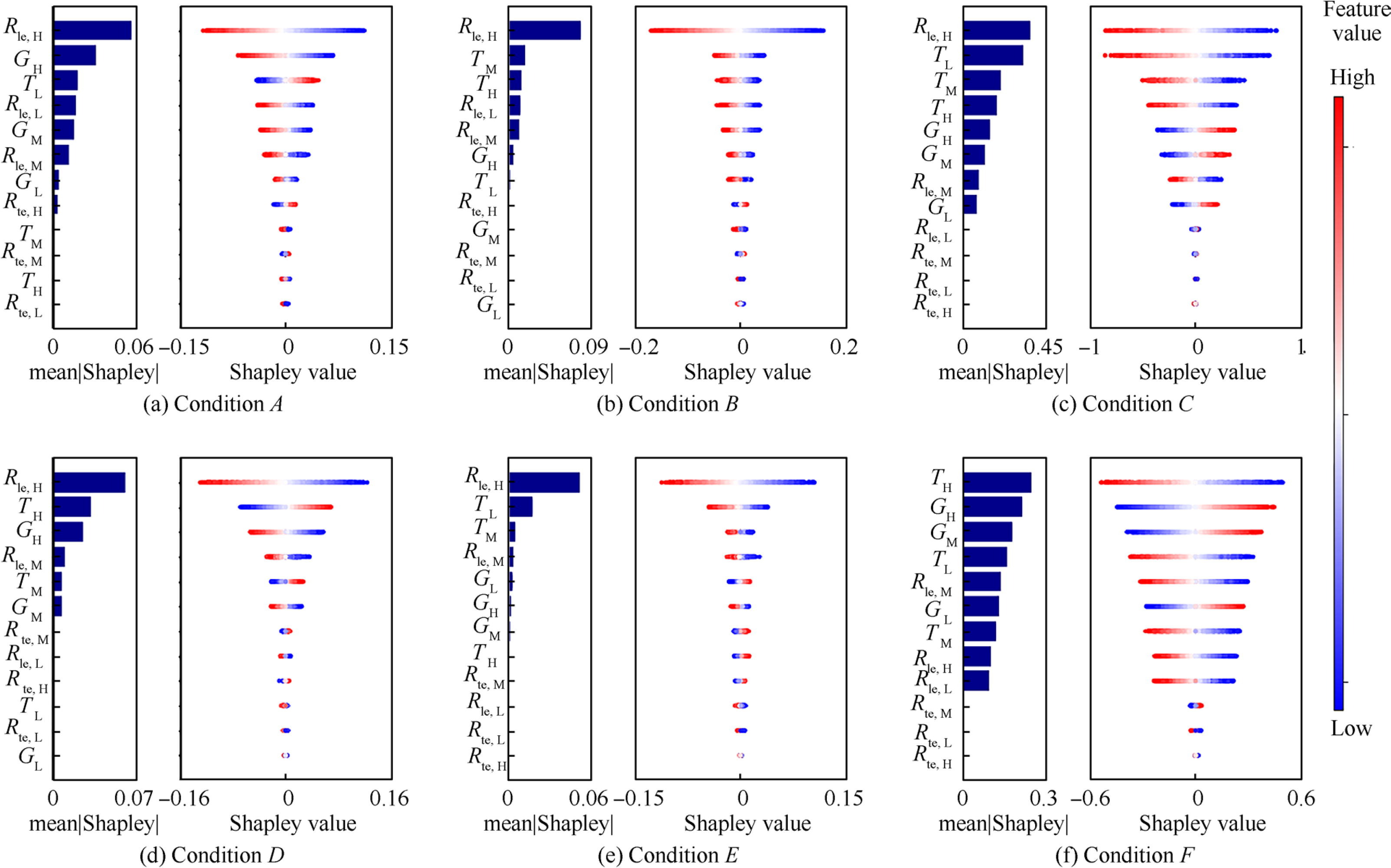}
    \caption{SHAP contribution of the geometrical uncertainties in the performance of the compressor (pressure ratio and efficiency) for six working conditions: near stall (A), peak efficiency (B), near choke (C) points for the supersonic/transonic flow and the same conditions (D, E, F) for subsonic flows. Figure extracted from \citet{junying2023} with permission from the publisher (Chinese Society of Aeronautics and Astronautics).}
    \label{fig:figure16}
\end{figure*}

The importance that rotating machines have in the performance of aerospace and industrial systems highlights the relevance of optimizing their designs and understanding the effect that the uncertainties~\citep{rezaeiravesh2018} have on the results. 

\subsubsection*{Explanation of complex flows}

The additive-feature-attribution methods have also been used for explaining complex flows involving fluid-particle interaction or multi-phase flows. For instance, the work of \citet{ouyang2023} focused on the first idea. A machine-learning methodology is used to link the mesoscale drag correction in fluid-particle flows with the physical parameters (velocity, pressure, particle volume fraction, and their gradients). This mesoscale drag correction is crucial for the sub-grid modeling required for the simulation of industrial scale flows~\citep{igci2008}. The interpretation of the markers on the mesoscale drag is presented from a data perspective using SHAP values to quantify their effects. In addition, due to the importance of particle conveying in industries such as mining or petroleum transportation~\citep{silva2015,ma2017}, other authors as \citet{xiao2024} have applied the SHAP values to link the pipe geometrical design, the conveying velocity as well as the particle concentration and diameter with the pressure drop and the erosion rate. This analysis showed that the conveying velocity is the most influential feature for the pressure drop, and the particle diameter, the conveying velocity, and the particle concentration for the erosion rate.

On the other hand, two-phase flows are present in a wide range of industrial applications such as chemical reactors, refrigeration processes, or nuclear power plants. The determination of the gas and liquid phases in a pipeline depends on the pipeline geometry and orientation relative to the gravity~\citep{taitel1990}. The identification of the flow regime can be addressed through visual inspection (requiring transparent pipe sections)~\citep{thaker2015} or using quantitative means measuring variations in the pressure or void fraction~\citep{ge2011,shaban2014}. The use of machine-learning techniques offers the possibility to identify the flow regime more objectively~\citep{cai1994}. For instance, in the work of \citet{khan2023} the flow regime is identified by using dynamic pressure signals. Then, the SHAP values are used to determine which input features are responsible for the identification of a particular flow regime.

Due to their capability of for detecting the relationships between the inputs and the outputs of the machine-learning models, the SHAP values have been widely applied to complex flows in which the analysis is complicated and requires machine-learning models for predicting the desired outputs.

\section{Conclusions and outlook}

Due to the complexity of studying turbulent flows, data-driven methods based on deep-learning models have been widely used for improving the results of numerical simulations, experimental tests, and optimization problems. The growing use of machine-learning models in fluid dynamics evidences the necessity of understanding the complex relationships and patterns detected in the training process.

In this manuscript, the additive-feature-attribution methods used in fluid mechanics are summarized, due to their linearity and easiness of interpretation. The SHAP values are described as the only possible solution to satisfy the properties required for providing a single solution for the interpretation of the model, as they satisfy the local accuracy, missingness, and consistency properties. Four different algorithms for simplifying the calculation of the attributions of the features are presented: kernel SHAP, tree SHAP, gradient SHAP, and deep SHAP.

The application of the previous methodologies in fluid-mechanics problems can be classified into three different main applications. The first group uses the additive-feature-attribution methods for explaining the RANS-model corrections calculated with data-driven methodologies. They link the physical properties of the flow used for modeling the eddy viscosity with their relevance in the model. Then, importance fields can be mapped for every physical variable to identify and understand which magnitudes are more important in each region of the flow. The second application is the use of the additive-feature-attribution methods for explaining the fundamentals of the turbulent phenomena. The SHAP values are calculated to rank the importance of the different structures or modes of the flow. Finally, other applications in fluid dynamics are mentioned, such as the understanding of the importance of the factors affecting the turbulent intensity near the airports or the aerodynamic load on a building surface.

Therefore, additive-feature-attribution methods have been demonstrated to be a relevant tool for interpreting the deep-learning models used in the fluid-mechanics field. SHAP values compute the importance of every single input feature, improving the connection between the results of the models and the physical phenomena, and providing a clear quantification of the input impact. 

Additive-feature-attribution methods provide a clear and unequivocal interpretation of the effects that an input produces on the output of a model. Using these methods for the analysis of turbulent flows provides deeper insight into an unsolved problem of classical physics with tremendous industrial and environmental implications.

Based on the examples provided in the present article, we believe that the use of SHAP values in turbulence research, and more generally in fluid mechanics and heat transfer, will provide new relevant directions of study in a wide range of applications. This additional insight into these phenomena has the potential of increasing the energetic efficiency of all those systems working with fluids. Furthermore, the SHAP framework may become a useful tool for facing global environmental problems in the near future.

\section*{Acknowledgments}
\label{sec:acknowledges}
RV acknowledges the financial support from ERC grant no. 2021-CoG-101043998, DEEPCONTROL. Views and opinions expressed are however those of the author(s) only and do not necessarily reflect those of the European Union or the European Research Council. Neither the European Union nor the granting authority can be held responsible for them. SH is grateful for the grant PID2021-128676OB-I00 funded by MCIN/AEI/10.13039/ 501100011033 and by “ERDF A Way of Making Europe”, by the European Union

%The data has been obtained with support of grant PID2021-128676OB-I00 funded by MCIN/AEI/10.13039/ 501100011033 and by “ERDF A way of making Europe”, by the European Union (SH).
 
% Use the style of numbering in square brackets.
% If nothing is used, default style will be taken.
%\begin{enumerate}[a)]
%\item 
%\item 
%\item 
%\end{enumerate}  

%\begin{itemize}
%\item 
%\item 
%\item 
%\end{itemize}  

%\begin{description}
%\item[]
%\item[] 
%\item[] 
%\end{description}  

%\clearpage %%Remove this from your manuscript

% Figure
%\begin{figure}%[]
%  \centering
%    \includegraphics{}
%    \caption{}\label{fig1}
%\end{figure}

%\begin{table}%[]
%\caption{}\label{tbl1}
%\begin{tabular*}{\tblwidth}{@{}LL@{}}
%\toprule
%  &  \\ % Table header row
%\midrule
% & \\
% & \\
% & \\
% & \\
%\bottomrule
%\end{tabular*}
%\end{table}

% Uncomment and use as the case may be
%\begin{theorem} 
%\end{theorem}

% Uncomment and use as the case may be
%\begin{lemma} 
%\end{lemma}


%\section{}\label{}

\printcredits

\bibliographystyle{cas-model2-names}

\bibliography{Review_XAI_v1}



\end{document}